\documentclass{emulateapj}

\usepackage{amsmath}

\newcommand{\be}{\begin{equation}}
\newcommand{\ee}{\end{equation}}
\newcommand{\cA}{{\cal A}}
\newcommand{\beps}{{\mbox{\boldmath $\epsilon$}}}
\newcommand{\bmu}{{\mbox{\boldmath $\mu$}}}
\newcommand{\bB}{{\bf B}}
\newcommand{\bE}{{\bf E}}
\newcommand{\bD}{{\bf D}}
\newcommand{\bH}{{\bf H}}
\newcommand{\bfe}{{\bf e}}
\newcommand{\bk}{{\bf k}}

\newcommand{\haty}{{\bf\hat y}}
\newcommand{\hatz}{{\bf\hat z}}
\newcommand{\etal}{{et al.}}

\newcommand{\snj}{\sin\theta^{(t)}_j }
\newcommand{\cnj}{\cos\theta^{(t)}_j }
\newcommand{\snjs}{\sin^2\theta^{(t)}_j }
\newcommand{\cnjs}{\cos^2\theta^{(t)}_j }
\newcommand{\sni}{\sin\theta^{(i)} }
\newcommand{\cni}{\cos\theta^{(i)} }

\newcommand{\cnti}{\cos 2\theta^{(i)}}
\newcommand{\snis}{\sin^2\theta^{(i)} }

\newcommand{\snp}{\sin\varphi }
\newcommand{\cnp}{\cos\varphi }
\newcommand{\snb}{\sin\theta_{B} }
\newcommand{\cnb}{\cos\theta_{B} }
\newcommand{\sntb}{\sin 2 \theta_{B} }
\newcommand{\cntb}{\cos 2 \theta_{B} }
\newcommand{\snps}{\sin^2\varphi }
\newcommand{\cnps}{\cos^2\varphi }
\newcommand{\snbs}{\sin^2\theta_{B} }
\newcommand{\cnbs}{\cos^2\theta_{B} }
\newcommand{\sntp}{\sin 2 \varphi }
\newcommand{\cntp}{\cos 2 \varphi }
\newcommand{\snif}{\sin^4\theta^{(i)} }

\newcommand{\sn}{\sin\theta^{(t)}}
\newcommand{\cn}{\cos\theta^{(t)}}
\newcommand{\cnre}{\cos\theta^{(t)}_R}
\newcommand{\cnim}{\cos\theta^{(t)}_I}
\newcommand{\bfs}{{\mbox{\boldmath $s$}}}

\shorttitle{Radiation from Condensed Surface of Magnetic Neutron Stars}
\shortauthors{van Adelsberg \etal}

\begin{document}

\title{Radiation from Condensed Surface of Magnetic Neutron Stars}

\author{Matthew van Adelsberg\altaffilmark{1},
Dong Lai\altaffilmark{1}, 
Alexander Y. Potekhin\altaffilmark{2,3},
Phil Arras\altaffilmark{4}}
\altaffiltext{1}{Center for Radiophysics and Space Research, Department
of Astronomy, Cornell University, Ithaca, NY 14853; mvanadel@astro.cornell.edu, 
dong@astro.cornell.edu}
\altaffiltext{2}{Ioffe Physico-Technical Institute,
  Politekhnicheskaya 26, 194021 St.~Petersburg, Russia; palex@astro.ioffe.ru}
\altaffiltext{3}{Isaac Newton Institute of Chile, 
         St.~Petersburg Branch, Russia}
\altaffiltext{4}{Kavli Institute for Theoretical Physics, University of California 
	at Santa Barbara; arras@kitp.ucsb.edu}

\slugcomment{Received 2004 May 28; accepted 2005 April 11}

\begin{abstract}
Recent observations show that the thermal X-ray spectra of many 
isolated neutron stars are featureless and in some cases (e.g.,
RX~J1856.5$-$3754) well fit by a blackbody. Such a perfect 
blackbody spectrum is puzzling since radiative transport through typical
neutron star atmospheres causes noticeable deviation from blackbody.  
Previous studies have shown that in a strong magnetic field, the outermost
layer of the neutron star may be in a condensed solid or liquid form
because of the greatly enhanced cohesive energy of the condensed matter.
The critical temperature of condensation 
increases with the magnetic field strength, and can be as high as 
$10^6$~K (for Fe surface at $B\sim 10^{13}$~G or H surface at
$B\sim {\rm a~few}\times 10^{14}$~G). Thus the thermal 
radiation can directly emerge from the degenerate metallic condensed surface,
without going through a gaseous atmosphere. Here
we calculate the emission properties (spectrum and polarization)
of the condensed Fe and H surfaces of magnetic neutron stars in the 
regimes where such condensation may be possible. 
For a smooth condensed surface, the overall emission is reduced from the 
blackbody by less than a factor of 2. The spectrum exhibits 
modest deviation from blackbody across a wide energy range, 
and shows mild absorption features associated with the ion 
cyclotron frequency and the electron plasma frequency in the condensed matter. 
The roughness of the solid condensate (in the Fe case) 
tends to decrease the reflectivity of the surface, and make the 
emission spectrum even closer to blackbody. We discuss the implications 
of our results for observations of dim, isolated neutron stars and magnetars.
\end{abstract}

\slugcomment{The Astrophysical Journal, 
scheduled to vol.\ 628, no.\ 2, 2005 July 20}


\section{Introduction}

In the last few years, much progress has been made in studying surface
radiation from isolated neutron stars (NSs) 
\citep[see, e.g.,][for a review]{PZ03}. 
So far about 20 NSs have been detected in thermal emission.
\setcounter{footnote}{4}
With the exception of 3-4 sources,\footnote{Spectral 
features have been detected in 1E 1207.4-5209 
\citep{sanwaletal02,mereghettietal02,haileymori02,DeLuca-ea},
RX J1308.6+2127 \citep{Haberl-ea03}, 
RX J1605.3+3249 \citep{Kerkwijk04} 
and possibly 
RX J0720.4$-$3125 \citep{Haberl-ea04a}. 
See also \citet{morihailey03} and \citet{HoLai04} 
for possible identifications of these features.}
the thermal spectra of most observed isolated NSs are featureless and 
sometimes well fit by a blackbody. 
For example, deep observations with
\textit{Chandra} and \textit{XMM-Newton} show that the soft X-ray (0.15-1~keV)
spectrum of RX J1856.5$-$3754 \citep{Walter-ea} can be fit with an almost perfect blackbody
at $kT=64$~eV \citep[e.g.,][]{Drake-ea,Burwitz-ea03}. 
The optical data of RX J1856.5$-$3754 is well represented
by a Rayleigh-Jeans spectrum, but the observed flux is a factor of
7 higher than extrapolation from the X-ray blackbody 
\citep[see][]{ponsetal02}. 
Thus the spectrum of RX J1856.5$-$3754 
is best fit by a two-temperature
blackbody model. Using this model as well as the 
observational upper limit ($1.3\%$ at $2\sigma$) 
of X-ray pulsation \citep{Burwitz-ea03}, \citet{BrajeRomani} 
obtained
several constraints on the viewing geometry, mass-to-radius ratio,
and temperature distribution.
Another much-studied, dim, isolated NS, RX J0720.4$-$3125
also shows an X-ray spectrum well fit by a blackbody at $T\simeq 1$~MK
(\citealt{Paerels-ea01}; but see \citealt{Haberl-ea04a} 
for possible spectral features).

The featureless, and in some cases ``perfect'' blackbody spectra
observed in isolated NSs are puzzling. This is because a NS atmosphere, 
like any stellar atmosphere, is not a perfect blackbody emitter 
due to nongrey opacities: On the one hand, a heavy-element (e.g., Fe) 
atmosphere would produce many spectral lines in the X-ray band 
\citep[e.g.,][]{RajagopalRomani,ponsetal02} 
on the other hand, a light-element
(e.g., H or He) atmosphere would result in an appreciable hard tail relative 
to the blackbody \citep[e.g.,][]{Shibanov-ea}. 

One physical effect that may help explain the observations is vacuum
polarization. Recent work has shown
that for surface magnetic fields $B\ga 10^{14}$~G, strong-field vacuum
polarization can significantly affect radiative transfer in NS atmospheres,
leading to depression of the hard spectral tail and suppression of the
(cyclotron or atomic) absorption lines 
\citep{LaiHo02,LaiHo03a,HoLai03,HoLai04,Ho-ea03,lloyd03}.
Indeed, \citet{HoLai03}
suggested that the absence of lines in the observed thermal spectra of
several anomalous X-ray pulsars
(e.g., \citealt{juettetal02,tiengoetal02,moriietal03,pateletal03}) 
can be naturally explained by the vacuum polarization effect.
For the dim isolated NSs RX J1308.6+2127 \citep{Haberl-ea03}
and RX J1605.3+3249 \citep{Kerkwijk04}, the observed line features
are consistent with surface fields $\la 10^{14}$~G, at which 
vacuum polarization does not affect the emergent spectrum
\citep{HoLai04}. In the case of RX J1856.5$-$3754, 
if the NS has a magnetar-like surface magnetic field 
\citep[see][]{MoriRuderman} 
it may be possible to explain the almost 
perfect X-ray blackbody with an atmosphere.
However, theoretical models for 
low-temperature ($kT\simeq 60$~eV) magnetar atmospheres 
are currently not available because of uncertainties in 
treating atomic, molecular opacities and dense plasma effects
in such cool atmospheres (see \citealt{Ho-ea03,PC03,PC04}).

Recently, several groups have suggested that the spectrum of RX J1856.5$-$3754 
might be explained if the NS has a condensed surface with no atmosphere
above it 
\citep{Burwitz-ea01,Burwitz-ea03,MoriRuderman,Turolla-ea}.
The notion that an isolated magnetic NS has a 
condensed surface was first put forward in the 1970s 
\citep[see][]{Ruderman71,Flowers-ea}
although these early studies overestimated the cohesive energy of Fe
solid at $B\sim 10^{12}$~G. Revised calculations yielded a much smaller
cohesive energy
\citep{mueller84,Jones,Neuhauser-ea}
making condensation unlikely for most observed NSs.
\citet{LS97} studied the phase diagram of the H surface layer of
a NS and showed that for strong magnetic fields, if the star 
surface temperature is below a critical value (which is a function of 
the magnetic field strength), the atmosphere can undergo a phase
transition into a condensed state \citep[see also][]{Lai01}.
For $B\ga 10^{14}$~G, this may occur even for temperatures as 
high as $10^{6}$ K.  This raises the possibility that the thermal 
radiation is emitted directly from the metal surface of the NS.

The thermal emission from condensed Fe surface of magnetic NSs
was previously studied by \citet{Brinkmann}
\citep[see also][]{Itoh,LenzenTruemper}
and shown to produce a rough blackbody
with reduced emissivity and a spectral feature at the electron plasma energy.
For the temperatures and magnetic fields ($T\ga 10^7$~K and 
$B=10^{12-13}$G, appropriate for accreting X-ray pulsars) 
considered by Brinkmann, the Fe surface is not expected to be in the 
condensed state. However, at lower temperatures appropriate for dim,
isolated NSs, or for higher $B$ appropriate for magnetars,
condensation remains a possibility \citep[see][]{Lai01}.
 
In this paper, motivated by recent observations of dim isolated NSs, 
we calculate the emissivity of condensed Fe or H surface of 
magnetic NSs in the regime where we expect condensation might be 
possible. Our study goes beyond previous work \citep{Brinkmann,Turolla-ea}
in that we calculate both the 
spectrum and polarization of the emission, and provide a more accurate treatment 
of the dissipative effect and transmitted radiation.  In previous works, the ions 
have been treated as fixed; while the exact dielectric tensor of the condensed 
matter is currently unknown, we also consider the alternate limit of free ions 
(see \S 2.2).


Regardless how the effect of ions in the dielectric tensor is treated,
we find appreciable difference between our result and that of 
Turolla et al. We traced the difference to their neglect of the 
ion effect, and their ``one-mode'' treatment of the transmitted 
radiation in the low-energy regime (see \S 4.1). Some of our preliminary 
results were reported in \citet{ArrasLai}.

This paper is organized as follows. Section \ref{sect:CondSurf} 
summarizes the basic 
properties of the condensed matter in strong magnetic fields.
The method for calculating the emission from the surface is
outlined in \S\ref{sect:Method} and numerical results 
presented in \S\ref{sect:Results}.
We discuss the implications of our results to observations of
dim isolated NSs and magnetars in \S\ref{sect:Concl}.

\section{Condensed Surface of Magnetic Neutron Stars}
\label{sect:CondSurf}

\subsection{Condition for Condensation}
\label{sect:Condensation}

It is well known that strong magnetic fields can 
qualitatively change the properties of atoms, molecules and
condensed matter. For $B\gg B_0=Z^2 e^3 m_{e}^2 c /\hbar^3 =
2.35\,Z^2\times10^{9}$~G (where $Z$ is 
the nuclear charge number), the electrons in an atom are confined to
the ground Landau level, and the atom is elongated, with greatly enhanced
binding energy. Covalent bonding between atoms leads to linear molecular 
chains, and interactions between molecular chains can lead to the formation of
three-dimensional condensed matter
\citep[see][for a recent review]{Lai01}.

For H, the phase diagram under different conditions has been studied.  
\citet{LS97} showed 
that in strong magnetic fields,
there exists a critical temperature $T_{\rm crit}$ 
below which a phase transition from gaseous to condensed state occurs,
with $kT_{\rm crit}$ about $10\%$ of the cohesive energy of the 
condensed hydrogen. Thus, $T_{\rm crit}\sim 8\times 10^4,\,
5\times 10^5,\,10^6$~K for $B=10^{13},\,10^{14},\,5\times 10^{14}$~G
\citep{Lai01}. 
An analogous ``plasma phase transition'' was also obtained
in an alternative thermodynamic model for magnetized hydrogen plasma \citep{PCS}.
While this model is more restricted than \citet{LS97} 
in that it does not include long H$_n$ chains,
it treats more rigorously atomic motion across the strong $B$ field
and Coulomb plasma nonideality. In the Potekhin et al. model, the density
of phase separation is roughly the same as in \citet{LS97}
(see eq.~[\ref{rho_s}] below), but the critical temperature is
several times higher.
Thus there is probably a factor of a few uncertainty 
in $T_{\rm crit}$. However, there is no question that for $T\la T_{\rm crit}/2$,
the H surface of the NS is in the form of the condensed metallic state, 
with negligible vapor above it.

For heavy elements such as Fe, no such systematic characterization of the phase 
diagram has been performed. Calculations so far have
shown that at $10^{12}-10^{13}$~G, a linear chain is unbound relative to
individual atoms for $Z\ga 6$ 
\citep{Jones,Neuhauser-ea} -- contrary
to earlier expectations \citep{Flowers-ea}.\footnote{For 
sufficiently large $B$, when $B\gg 10^{14}(Z/26)^3$~G, we expect the 
linear chain to be bound in a manner similar to the H chain \citep{Lai01}.}
Therefore chain-chain interactions play a crucial role in determining whether
3D zero-pressure condensed matter is bound or not. Numerical
results of \citet{Jones}, together with approximate scaling relations suggest
an upper limit of the cohesive energy (for $Z\ga 10$)
$Q_s\la Z^{9/5}B_{12}^{2/5}$~eV, where $B_{12}=B/(10^{12}~{\rm G})$.
Thus for Fe, the critical temperature for phase transition $T_{\rm crit}\la
0.1Q_s/k\la 10^{5.5}B_{12}^{2/5}$~K \citep{Lai01}.  

The zero-pressure density of the condensed matter can be estimated as
\be
\rho_s \simeq 560\,\eta\,A\,Z^{-3/5}\,B^{6/5}_{12}~{\rm g~cm}^{-3},
\label{rho_s}
\ee
where $A$ is the mass number of the ion ($A\approx 1.007$ for H,
$A\approx 55.9$ for Fe), and 
$\eta=1$ corresponds to the uniform electron gas model in the
Wigner-Seitz approximation \citep{Kadomtsev}. Other effects 
(e.g., Coulomb exchange interaction, nonuniformity of the electron gas)
can reduce the density by up to a factor of $\sim 2$, and thus $\eta$ may be
as small as $0.5$ 
(\citealt{Lai01}; see also \citealt{PC04}). 
The condensate will be in the liquid state when the Coulomb coupling parameter
$\Gamma=(Ze)^2/(a_ikT) =0.227\,Z^2 (\rho_1/A)^{1/3} /T_6 < \Gamma_m$. 
Here, $a_i$ is the ion sphere radius ($(4\pi a_i^3/3)^{-1}=n_i$, where
 $n_i$ is the number density of ions),
 $\rho_1=\rho_s/(1\mbox{ g cm}^{-3})$,
$T_6=T/(10^6~{\rm K})$, and $\Gamma_m$ is the characteristic
value of $\Gamma$ at which the Coulomb crystal melts. In the one-component plasma model 
(i.e., classical ions on the background of the uniform degenerate electron gas),
$\Gamma_m = 175$, but the electron gas nonuniformity (i.e., electron
screening) introduces a dependence of  $\Gamma_m$
on $\rho$ and $Z$, typically within the range 
$\Gamma_m\sim160$--190 \citep{PC00}. From 
equation (\ref{rho_s}) we obtain $\Gamma\simeq1.876\,\eta^{1/3}Z^{9/5}B_{12}^{2/5}/T_6$
at the condensed surface. Therefore, the surface will be solid  
when $T < 7\times 10^4\eta^{1/3}(175/\Gamma_m)B_{14}^{2/5}$~K 
for H (where $B_{14}=B/10^{14}~{\rm G}$) 
and $T < 4\times 10^6\eta^{1/3}(175/\Gamma_m)B_{12}^{2/5}$~K for Fe.  
Therefore, if condensation occurs ($T < T_{\rm crit}$), we expect the Fe
condensate to be solid.
Note that we use the 
simple melting criterion above for the condensed phase only.  It cannot be used for 
non-condensed iron at $T\la 10^7$~K (e.g., when $T$ is only slightly above 
$T_{\rm crit}$)
because in this case the state of matter is affected by partial ionization.

\subsection{Dielectric Tensor of Condensed Matter}
\label{sect:DielTens}

The emissivity of the condensed NS surface will depend on its
(complex) dielectric tensor (see \S\ref{sect:Method}). As a first approximation, we 
consider the free electron gas model for the condensed matter
\citep[e.g.,][]{AshcroftMermin}. In the coordinate system with 
magnetic field $\bf B \rm$ along the $z$-axis, the dielectric tensor 
takes the form (cf. \citealt{Ginzburg})\footnote{See also \citealt{LaiHo03a}.  Note that eq.~(13) of \citealt{LaiHo03a} is incorrect: $\gamma_{ei}^{\pm}$ should 
simply be $\gamma_{ei}(1 + Z m_e/A m_p)$.  We neglect the factor 
$1 + Z m_e/A m_p$ in eq.~(\ref{eq:DieEle1}) since it provides a negligible 
correction relative to the uncertainty in the collisional damping (see \S 2.3).}
\begin{equation} \label{eq:PlaDie}
\bigl[\,\beps\,\bigr]_{{\bf \hat z} = {\bf \hat B} }\ 
= \left ( \begin{array}{ccc} \epsilon & i\,g & 0 \\ -i\,g & \epsilon & 0 \\ 0 & 0 & \eta \end{array} \right )
\end{equation} 
where 
\begin{subequations}
\label{eq:DieEle}
\begin{eqnarray}
\label{eq:DieEle1} \epsilon \pm g & \simeq & 1 - {{v_{e}} \over {(1\pm u_{e}^{1/2})(1\mp u_{i}^{1/2})+i\,\gamma_{ei}^{(tr)}}} \\
\label{eq:DieEle2} \eta & \simeq & 1 - {{v_{e}} \over {1 + i\,\gamma^{(l)}_{ei}}}.
\end{eqnarray}
\end{subequations}\hspace{-4pt}
In eq.~(\ref{eq:DieEle}), the dimensionless quantities $u_e=(E_{Be}/E)^2$, 
$u_i=(E_{Bi}/E)^2$, $v_e=(E_{pe}/E)^2$
are used, where $E=\hbar\omega$ is the photon energy,
$E_{Be}$, $E_{Bi}$ are the electron and ion cyclotron
energies, and $E_{pe}$ is the electron plasma energy.
These energies take the values:
\begin{subequations}
\label{eq:PEnergies}
\begin{eqnarray}
&& E_{Be}  =  {{\hbar e B} \over {m_{e} c}} = 1158\, B_{14} \mbox{ keV},\\
&& E_{Bi} =  {{\hbar Ze B} \over {m_{i} c}} = 0.635\, B_{14}\left ( {Z \over A} \right ) \mbox{ keV}, \\
&& E_{pe} = \left({{4\pi\hbar^2e^2 n_{e}} \over m_{e} } \right)^{1/2} 
= 0.0288\,\left ( {Z \over A} \right )^{1/2}\,\rho_{1}^{1/2}
\mbox{ keV} 
\nonumber\\&&\qquad
= 10.8\,\eta^{1/2} Z^{1/5}\,B_{14}^{3/5} \mbox{ keV},
\end{eqnarray}
\end{subequations}
\hspace{-4pt}where 
$n_{e}$ is the electron number density, and $m_{i}$ is the ion mass.
The collisional damping is calculated for motions transverse and longitudinal
with respect to the magnetic field.  
The dimensionless damping rates $\gamma^{(tr)}_{ei}$ and $\gamma^{(l)}_{ei}$ are 
obtained from the collisional damping rates $\nu_{ei}^{(tr)}$ and 
$\nu_{ei}^{(l)}$ (see \S\ref{sect:Damping}) through $\gamma_{ei}^{(tr)}=
\hbar \nu_{ei}^{(tr)}/E$ and 
$\gamma_{ei}^{(l)}=\hbar \nu_{ei}^{(l)}/E$.

Equations (\ref{eq:DieEle}) give the elements of the
dielectric tensor for a cold, magnetized plasma. While the
expressions were derived classically, the quantum
calculation, incorporating the quantized nature of
electron motion transverse to the magnetic field, yields
identical results
\citep[e.g.,][]{CanutoVentura,Pavlov-ea80}. More
significantly, expressions (\ref{eq:DieEle}) assume that
the electrons and ions are subject to the pairwise Coulomb
attraction, the interaction with the stationary magnetic
field, and the periodic force from the propagating
electromagnetic wave.  At high densities, however, other
interactions can also be important. For instance, the ions
are strongly coupled to each other when the Coulomb
parameter $\Gamma$ is large. It is this coupling that leads
to the liquid-solid phase transition mentioned in  \S2.1.
One might suggest that in the solid phase the ion motion
should be frozen (by setting the ion mass $m_i=\infty$), as
implicitly adopted by \citet{Turolla-ea}. This is not
exactly true. It is known that optical modes of a crystal
lattice (at $B=0$) can be described by polarizability of
the form given by equation~(\ref{eq:DieEle}) with an
additional term  in the denominator which specifies the
binding of the ions \citep[see, e.g.,][]{Ziman79}.
According to the harmonic model of the Coulomb crystal
\citep{Chabrier93}, the characteristic ion oscillation
frequency (the Debye frequency of acoustic phonons) is
$\omega_D \approx 0.4 E_{pi}/\hbar$, where $E_{pi}
=6.75\times 10^{-4}\,(Z/A)\,\rho_1^{1/2}$ 
keV is the ion plasma energy. 
The magnetic field appreciably affects the motion of the ions 
in the Coulomb crystal if $\hbar\omega_D/E_{Bi}\lesssim 1$ 
(or $E_{pi}\lesssim E_{Bi}$, see \citealt{Usov-ea80}).
From equations
(\ref{eq:PEnergies}) we find  $\hbar\omega_D/E_{Bi}\approx
1.6\eta^{1/2} A^{1/2} Z^{-0.3} B_{14}^{-0.4},$  which shows
that the magnetic forces on the ions are not completely
negligible compared to the Coulomb lattice forces.

Needless to say, our current understanding of the condensed
matter in strong magnetic fields is crude, and equations
(\ref{eq:DieEle}) are only a first approximation to the true
dielectric tensor of the magnetized medium. In our
calculations below, in addition to the case of of
quasi-free ions described by equations~(\ref{eq:DieEle}),
we shall also consider the 
case where the motion of
the ions is neglected (formally obtained by setting
$m_i=\infty$). It is reasonable to expect that in reality the
surface radiation spectra lie between the results obtained
for these two limiting cases.  
Nevertheless, future work is needed 
to evaluate the reliability of our results at low frequencies.


\subsection{Collisional Damping Rate in the Condensed Matter}
\label{sect:Damping}

\begin{figure}
\epsscale{1.0}
\plotone{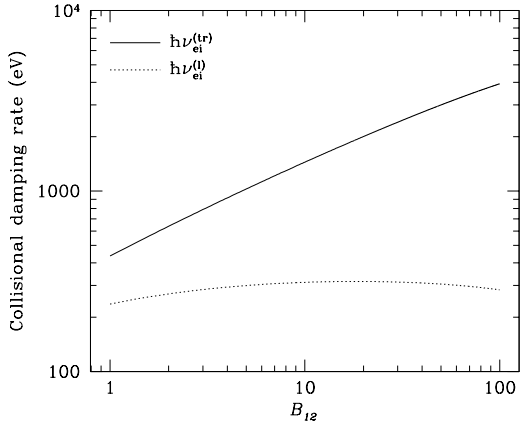}
\caption{Transverse and longitudinal damping rates $\hbar \nu_{ei}^{(tr)}$ and 
$\hbar \nu_{ei}^{(l)}$ as a function of magnetic field strength $B=10^{12} B_{12}$ for condensed 
Fe surface at $T=10^6$ K.  
The density is calculated using eq.~(\ref{rho_s}), with 
$\eta=1$. 
}
\label{fig1}
\end{figure}

For the collisional damping rates $\gamma^{(l,tr)}_{ei}$,
different approximations can be used in different ranges of
frequency $\omega$ and density $\rho$.
For $E\gg E_{pe} \equiv \hbar\omega_{pe}$ (where $\omega_{pe}$
is the electron plasma frequency),
the electron-ion collisions can be considered as independent, and
$\gamma^{(l,tr)}_{ei}$ are determined by the effective
rates of free-free transitions of a single electron-ion pair.
However, this approximation fails at $E\lesssim E_{pe}$, where
collective effects become important.
Moreover, the electron degeneracy should be taken into account
in the condensed surface.
In general, the complex dielectric tensor $\beps$ for arbitrary 
$\omega$ can be obtained from kinetic theory, at least in principle
\citep[e.g., see][]{Ginzburg}. Since such a general expression of
$\beps$ is unknown at present, we shall approximate $\gamma^{(l,tr)}_{ei}$ 
in the $E\lesssim E_{pe}$ regime using the result of 
Potekhin (1999), who obtained the zero-frequency
conductivity tensor for degenerate Coulomb plasmas (liquid and solid)
in arbitrary magnetic fields. Specifically, we set 
$\nu^{(l)}_{ei}=1/\tau_\|,~\nu^{(tr)}_{ei}=1/\tau_\perp$,
where $\tau_\|$ and $\tau_\perp$ are the effective collision times
given by eqs. (28) and (39) of \citet{P99}, respectively.
Figure \ref{fig1} shows $\hbar \nu_{ei}^{(tr)}$ and $\hbar \nu_{ei}^{(l)}$ 
as a function of magnetic field strength
for condensed Fe surface at $T=10^6$~K, over the range 
$B=10^{12}-10^{14}$~G.

The calculations of $\nu^{(l)}_{ei}$ and $\nu^{(tr)}_{ei}$ 
adopted in our paper neglect the influence of the magnetic field 
on the motion of the ions.  Therefore, these calculations apply only 
in the $u_i\rightarrow 0$ limit (this corresponds to the ``fixed'' 
ion limit of \S 2.2), or in the regime $E\gg E_{Bi}$.
We note, however, that the emissivity at $E\la E_{Bi}$
does not depend sensitively on the damping rates
(see \S 4; in particular, Fig.~2 shows that the emissivity at such 
low energies is almost the same with/without damping). Thus, unless
the true values of $\nu^{(l)}_{ei},~\nu^{(tr)}_{ei}$ at such low energies
are many orders of magnitude larger than our adopted values,
our emissivity results will not be affected by this uncertainty
(Indeed, as discussed in \S 2.2, the main uncertainty at such low energies
lies in whether to treat the ions as ``free'' or ``fixed'').

\section{Emission From Condensed Matter: Method}
\label{sect:Method}

In this paper we consider the regime where a clear phase separation occurs at the NS surface 
(i.e., for $T$ at least a few times lower than $T_{\rm crit}$), so that the vapor (gas) above 
the condensed surface has negligible density and optical depth.  In this case the radiation 
emerges directly from the condensed matter.
\subsection{Kirchhoff's Law For A Macroscopic Object}

A macroscopic body at temperature $T$ produces an intrinsic thermal emission,
with specific intensity $I_\nu^{(e)}$.  To calculate the 
intensity, consider the body placed inside a blackbody cavity also 
at temperature $T$, i.e., the body is in thermodynamical
equilibrium with the surrounding radiation field, whose intensity
is given by the Planck function $B_\nu(T)$.
Imagine a ray of the cavity radiation impinging on a surface 
element $dA$ of the body. The radiation field is unpolarized, and the 
electric field of the incoming ray can be written in 
terms of two independent polarization states: $\bf E\rm^{(i)}_{1}
=\cA\bf e\rm^{(i)}_{1}$ and $\bf E\rm^{(i)}_{2}=\cA\bf e\rm^{(i)}_{2}$, 
where $\cA=\sqrt{B_{\nu}/2}$, and ${\bf e}^{(i)}_1$ and 
${\bf e}^{(i)}_2$ are the polarization eigenvectors of the incident wave.
The ray is, in general, partially reflected, each incoming 
polarization giving rise to a reflected field:
\begin{subequations}
\label{eq:Eref}
\begin{eqnarray}
\label{eq:Eref1}
&&{\bf E}^{(r)}_{1}=\cA\left(r_{11}{\bf e}^{(r)}_{1}+r_{12}{\bf
e}^{(r)}_{2}\right), \\
\label{eq:Eref2}
&&{\bf E}^{(r)}_{2}=\cA\left(r_{21}{\bf e}^{(r)}_{1}+r_{22}{\bf
e}^{(r)}_{2}\right),
\end{eqnarray}
\end{subequations}\hspace{-4pt}
where $\bf E\rm^{(r)}_{1}$ and $\bf E\rm^{(r)}_{2}$ are the reflected 
electric fields due to incoming fields $\bf E\rm^{(i)}_{1}$ 
and $\bf E\rm^{(i)}_{2}$, respectively.  Thus, the intensity of 
radiation in the reflected field with polarizations $\bf e\rm^{(r)}_{1}$ 
and $\bf e\rm^{(r)}_{2}$ is:
\begin{subequations}
\begin{eqnarray}
&& I^{(r)}_{\nu 1}={1\over 2}\left(\left|r_{11}\right|^2+\left|r_{21}\right|^2
\right)B_{\nu}\equiv {1\over 2}R_1B_\nu, \\
&& I^{(r)}_{\nu 2}={1\over 2}\left(\left|r_{12}\right|^2+\left|r_{22}\right|^2
\right)B_{\nu}\equiv {1\over 2}R_2B_\nu.
\end{eqnarray}
\end{subequations}\hspace{-4pt}
The energy in the incoming wave for frequency band $\nu\rightarrow \nu+d\nu$ 
during time $dt$ is $B_{\nu}\,dA\,d\Omega^{(i)}\,d\nu\,dt$, 
where $d\Omega^{(i)}$ is the solid angle element around the direction of the 
incoming ray.  Similarly, the energy contained in the reflected wave 
(along each polarization) is 
$(1/2)R_{1,2}\,B_{\nu}\,dA\,d\Omega^{(r)}\,d\nu\,dt$, with
$d\Omega^{(r)}=d\Omega^{(i)}$. To insure that the cavity radiation field
remains an unpolarized blackbody, the intensities 
of radiation emitted by the body (in the same direction as the reflected
wave) with polarizations $\bf e\rm^{(r)}_{1}$ and $\bf e\rm^{(r)}_{2}$ must be:
\begin{subequations}
\label{eq:Spectrum}
\begin{eqnarray}
&&I^{(e)}_{\nu 1}={1\over 2}B_{\nu}-I^{(r)}_{\nu 1}={1\over 2}(1-R_1)B_\nu, \\
&&I^{(e)}_{\nu 2}={1\over 2}B_{\nu}-I^{(r)}_{\nu 2}={1\over 2}(1-R_2)B_\nu.
\end{eqnarray}
\end{subequations}\hspace{-4pt}
Since $I^{(e)}_{\nu 1}$ and $I^{(e)}_{\nu 2}$ are intrinsic properties
of the body, these equations should also apply even when the body is not 
in thermodynamical equilibrium with a blackbody radiation field.
Thus, a body at temperature $T$ has emission intensity
\be
\label{eq:TotSpec}
I_\nu^{(e)}=(1-R)B_\nu(T) \equiv J B_{\nu}(T)
\ee
where $R\equiv (1/2)(R_1+R_2)$ is the reflectivity, and $J = 1-R$ is the dimensionless emissivity.
The degree of linear polarization of the emitted radiation is 
\be 
\label{eq:EmitPol}
P\equiv {{I^{(e)}_{\nu 1}-I^{(e)}_{\nu 2}}\over{I^{(e)}_{\nu 1}+I^{(e)}_{\nu2}}}
={1\over 2}{{R_{2}-R_{1}}\over {1-R}}.
\ee


\subsection{Calculation of Reflectivity}
\label{sect:Reflect}

To calculate the reflectivity $R$, we set up a coordinate system as follows:
the surface lies in the $xy$ plane with the $z$-axis along the surface
normal. The external magnetic field $\bB$ lies in the $xz$ plane,
with $\hat\bB\times\hatz=\sin\theta_B\haty$, where $\theta_B$ is the angle
between $\hat\bB$ and $\hatz$. Consider a ray (of certain polarization,
${\bf e}_1^{(i)}$ or ${\bf e}_2^{(i)}$) impinging on the surface,
with incident angle $\theta^{(i)}$ and azimuthal angle $\varphi$, so that
the unit wave vector $\hat\bk^{(i)}=(-\sin\theta^{(i)}\cos\varphi,
-\sin\theta^{(i)}\sin\varphi,-\cos\theta^{(i)})$. The transmitted (refracted)
and reflected rays lie in the same plane as the incident ray.
Our goal is to calculate the field associated with the reflected ray.

Outside the condensed medium ($z>0$), the dielectric tensor
and permeability tensor are determined by the vacuum polarization effect
with $\beps=a\mbox{\boldmath $I$}+q\hat\bB\hat\bB$ and $\bmu^{-1}=a\mbox{\boldmath $I$}+m\hat\bB\hat\bB$,
where $a,q,m$ are dimensionless
functions of $B$ \citep[see][and references therein]{HoLai03}.
Since $a\sim 1$ and $|q|,~|m|\ll 1$ for $B\ll 5\times 10^{16}$~G, 
the vacuum polarization effect is negligible. In our calculation
(Appendix A), we choose ${\bf e}^{(i)}_1$ (and ${\bf e}^{(r)}_1$)
to be along the incident plane and ${\bf e}^{(i)}_2$ 
(and ${\bf e}^{(r)}_2$) perpendicular to it.

Consider an incident ray with $\bE^{(i)}=
\bE_1^{(i)}=\cA{\bf e}_1^{(i)}$. The 
$\bE$-field of the reflected ray takes the form given by eq.~(\ref{eq:Eref1}),
while the transmitted wave has the form:
\be
\bE^{(t)}=\bE_1^{(t)}=\cA\,\left(t_{11}\bfe_1^{(t)}+t_{12}\bfe_2^{(t)}
\right).
\ee
The eigenvectors of the transmitted wave, $\bfe_1^{(t)}$ and $\bfe_2^{(t)}$,
depend on the refraction angles, $\theta_1^{(t)}$ and $\theta_2^{(t)}$,
respectively; note that in general, these angles are complex and
different from each other. The refraction angle $\theta_j^{(t)}$,
the mode eigenvector $\bfe_j^{(t)}$ and the corresponding index of refraction 
$n_j^{(t)}$ ($j=1,2$) satisfy Snell's law
\be
\label{eq:Snell}
\sin\theta^{(i)}=n_j^{(t)}\sin\theta_j^{(t)},
\ee
and the mode equation\footnote{The vacuum polarization effect is neglected 
in eq.~(\ref{eq:ModeEq}), which is justified because
the density of the condensed medium is much larger than the vacuum resonance density, 
$\rho_V\simeq 0.96 (A/Z) B_{14}^2 (E/\mathrm{keV})^2 \mbox{ g cm}^{-3}$ 
\citep[see][]{LaiHo02}.}
\be
\label{eq:ModeEq}
\left[\beps+(n_j^{(t)})^2
\left(\hat\bk^{(t)}_j\hat\bk^{(t)}_j-{\bf I}\right)\right]\cdot\bE_j^{(t)}=0,
\ee
where ${\bf I}$ is the unit tensor,
and $\hat\bk^{(t)}_j=(-\sin\theta_j^{(t)}\cos\varphi, 
-\sin\theta_j^{(t)}\sin\varphi,-\cos\theta_j^{(t)})$ is the unit wavevector of the transmitted waves.

In the xyz coordinate system, the rotated dielectric tensor takes the form:
\be
\label{eq:RotDie}
\!\bigl[\,\beps\,\bigr] \!\! = \!\!\! \left(\!\!\begin{array}{ccc}
\epsilon\cnbs\!\!+\!\eta\snbs & ig\cnb & (\epsilon\!-\!\eta)\!\snb\cnb\\
-ig\cnb & \epsilon & -ig\snb\\
(\epsilon-\eta)\snb\cnb & ig\snb & \epsilon\snbs+\eta\cnbs
\end{array}\!\!\!\right)
\ee
For eq.~(\ref{eq:ModeEq}) to have a non-trivial solution, the determinant of the matrix
$\beps+(n_j^{(t)})^2\left(\hat\bk^{(t)}_j\hat\bk^{(t)}_j-{\bf I}\right)$
must be equal to 
zero.  This gives an equation involving powers of $n_j^{(t)}$, $\snj$, and $\cnj$.  
Substituting eq.~(\ref{eq:Snell}) into this equation, and squaring both sides yields a fourth-order 
polynomial in $(n_j^{(t)})^2$, which allows for the determination of the indices of 
refraction (see Appendix \ref{sect:A} for more details).  Having determined $n_j^{(t)}$, eq.~(\ref{eq:ModeEq}) 
can be used to calculate $\bfe_j^{(t)}$, while eq.~(\ref{eq:Snell}) gives $\theta_j^{(t)}$.
Once $\theta_j^{(t)}$, $\bfe_j^{(t)}$ and $n_j^{(t)}$ are known,
$r_{11},\,r_{12},\,t_{11}$ and $t_{12}$ can be obtained
using the standard electromagnetic boundary conditions:
\be
\label{eq:BCs}
\Delta\bD\cdot\hatz=0,\quad
\Delta\bB\cdot\hatz=0,\quad
\Delta\bE\times\hatz=0,\quad
\Delta\bH\times\hatz=0,
\ee
where, e.g., $\Delta\bE\equiv \bE^{(i)}+\bE^{(r)}-\bE^{(t)}$, ${\bf D}^{(t)} = \beps\cdot \bE^{(t)}$, and
\be
\bH^{(t)}={\bf B}^{(t)} = \cA \left(
n_1^{(t)} t_{11}\,{\bf\hat k}^{(t)}_1\times {\bf e}^{(t)}_1
+n_2^{(t)} t_{12}\,{\bf\hat k}^{(t)}_2\times {\bf e}^{(t)}_2
\right),
\ee
neglecting the vacuum polarization effect ($\bmu\simeq {\bf I}$).  Note that eqs.~(\ref{eq:BCs}) 
are not all independent.  Only $\Delta\bE\times\hatz=0$ and 
$\Delta\bB\times\hatz=0$ are used in our calculation.

A similar procedure applies in the case when the incident wave is
$\bE^{(i)}=\bE_2^{(i)}=\cA\bfe_2^{(i)}$, yielding the reflection coefficients
$r_{21}$ and $r_{22}$ (together with $t_{21},\,t_{22}$).

\section{Emission from Condensed Surface: Results}
\label{sect:Results}

\begin{figure}
\epsscale{1.0}
\plotone{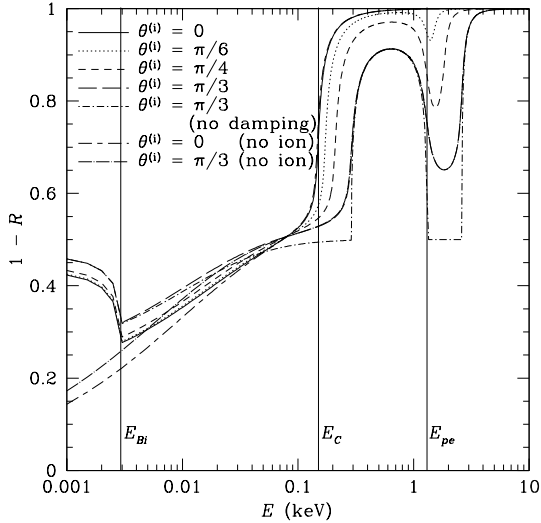}
\caption{Dimensionless emissivity $J=1-R$ as a function of photon energy 
$E$ (keV) for the case of condensed Fe surface, at 
$B=10^{12}$ G.  The $B$ field is normal to the surface.  
The different curves correspond to different 
angles $\theta^{(i)}$ between incident photon direction and surface 
normal.  The short-dashed-dotted line (labeled ``no damping'')
shows the result when the collisional damping is 
set to zero in the plasma dielectric tensor.   The other light lines 
(labeled ``no ion'') show 
results when ion motion is neglected for two values of $\theta^{(i)}$ (i.e., by 
setting the ion mass to $\infty$; see \S 2.2).
The three vertical lines denote the ion cyclotron energy $E_{Bi}$,
the electron plasma energy $E_{pe}$ [see eq.~(\ref{eq:PEnergies})]
and $E_c$ [eq.~(\ref{eq:ec})].} 
\label{fig2}
\end{figure}

\begin{figure}
\epsscale{1.0}
\plotone{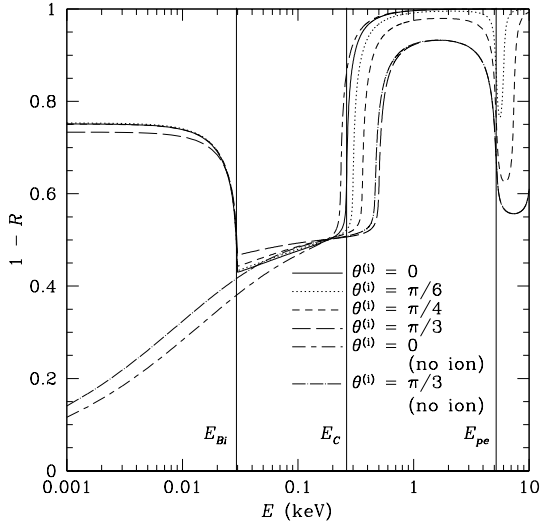}
\caption{Same as Fig.~2, except for $B=10^{13}$~G.}
\label{fig3}
\end{figure}

\begin{figure}
\epsscale{1.0}
\plotone{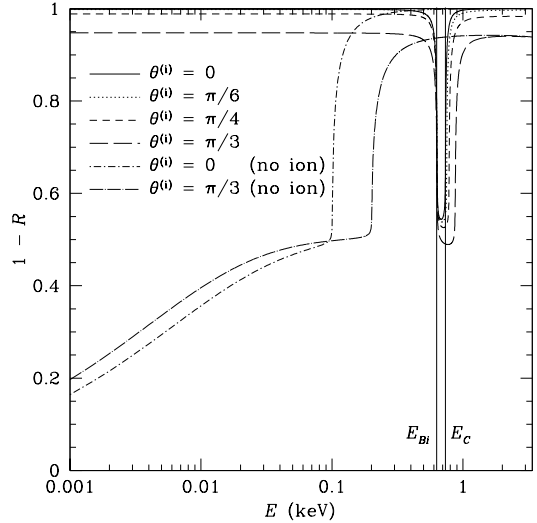}
\caption{Same as Fig.~2, except for H surface at $10^{14}$~G.}
\label{fig4}
\end{figure}

In this section, we present results of surface emission for three illustrative cases:
Fe surface at $B=10^{12}$~G and $10^{13}$~G, and H surface at $10^{14}$~G. As discussed 
in \S\ref{sect:Condensation}, the condensation temperature for these cases is around $10^6$~K.
Note that the dimensionless 
emissivity $J=1-R$ [see eq.~(\ref{eq:TotSpec})] depends only weakly on $T$ through the collisional 
damping rate (\S\ref{sect:DielTens}).  For concreteness, we set $T = 10^6$ K in all 
our calculations.  Figures \ref{fig2}--\ref{fig4}
show the emissivity $J$ as a function of photon energy $E$ in the three cases; the $B$ field 
is assumed to be normal to the surface ($\theta_B = 0$).
In all three cases, the emissivity is reduced from blackbody 
at low energies, while approaching unity for $E>{\rm a~few}\times E_{pe}$.  
In the case of Fe, there are features associated with 
the ion cyclotron energy $E_{Bi}$ and the electron plasma energy
$E_{pe}$.  For H, the electron plasma 
energy is too high to be of interest for observation, but the feature 
around the ion cyclotron energy is evident.

The spectral feature in the emissivity $J$ near $E_{Bi}$ can be understood by considering the special case of $\theta^{(i)}=0$ (normal incidence).  In this case the reflectivity takes the analytic form:
\begin{equation} \label{eq:NorInR}
R={1\over 2}\left|{{n_1-1}\over {n_1+1}}\right|^2+{1\over 2}\left|{{n_2-1}\over {n_2+1}}\right|^2,
\end{equation}
where $n_1$ and $n_2$ are the indices of refraction of the two modes in the medium,  
and are given by $n_{1}^2=\epsilon+g$, $n_{2}^2=\epsilon-g$.  Consider 
energies around $E_{Bi}$, such that $v_{e}, u_{e} \gg 1$.  We find
\begin{eqnarray}
\label{eq:AprInd}
n_{1,2}^2 &\approx& 1\mp {v_{e}(1\mp u_i^{1/2})\over u_{e}^{1/2}(1\mp u_i^{1/2})^2+
(\gamma_{ei}^{(tr)})^2}
\nonumber\\&&
+i{v_e\gamma_{ei}^{(tr)}\over u_e(1\mp u_i^{1/2})^2+
(\gamma_{ei}^{(tr)})^2}.
\end{eqnarray}
Although $\gamma_{ei}^{(tr)}$ can be greater than unity (see Fig.~\ref{fig1}), the 
imaginary part of $n_{1,2}^2$ can be neglected for a qualitative understanding of 
the spectral features, since $v_e/u_e \ll 1$ [see eq.~(\ref{eq:PEnergies})].  Then 
eq.~(\ref{eq:AprInd}) becomes
\be
n_{1,2}^2\approx 1\mp {v_{e}\over u_{e}^{1/2}(1\mp u_i^{1/2})}
\ee
For $E < E_{Bi}$ ($u_{i} > 1$), both $n_1$ and $n_2$ are real and differ from unity, 
leading to $J<1$.  
For $E > E_{Bi}$, $n_1$ is imaginary until
$(v_e/u_e^{1/2})(1-u_i^{1/2})^{-1}<1$, which occurs at
\be 
E_{C}\approx E_{Bi} + {E_{pe}^2 / E_{Be}}.
\label{eq:ec}\ee
Thus, for $E_{Bi} < E < E_{C}$, $n_1^2$ increases from $-\infty$ to $0$ 
(implying no mode propagation in the medium), giving rise to the broad depression 
in $J$ (with $J\rightarrow 0.5$ as the energy nears $E_{C}$).

We can similarly understand the feature near the electron plasma energy.  This feature 
appears only for $\theta^{(i)}\ne 0$. 
For energies around $E_{pe}$, $u_e\gg 1$, $u_i\ll 1$, and
we have $\epsilon\approx 1+v_e/u_e$ and $g\approx -v_e/u_e^{1/2}$. Substituting 
these values into
($\ref{eq:ModeEq}$) and neglecting terms to order $v_e/u_e$ and higher, we find
\begin{equation} \label{eq:ApproxInd}
n_1^2\approx 1+{v_e\over {(1-v_e)}}\snis,\quad n_2^2\approx 1
\end{equation}
For $E> E_{pe}$, both $n_1$ and $n_2$ are real, while for $E<E_{pe}$, 
$n_1^2<0$. The reflectivity no longer takes
the simple analytic form of ($\ref{eq:NorInR}$). However, the basic behavior of
the reflectivity is largely
the same: for one mode with imaginary $n$ and the
other with $n\approx 1$, the emissivity $J$ attains a local minimum ($\simeq 0.5$
in the absence of collisional damping; see Fig.~\ref{fig2}).

\begin{figure}
\epsscale{1.0}
\plotone{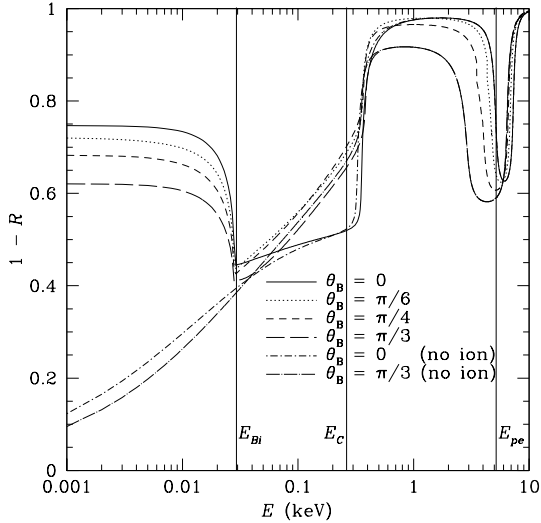}
\caption{Dimensionless emissivity $J=1-R$ 
as a function of photon energy $E$ 
for the case of condensed Fe surface at $B=10^{13}$ G.  The incident angles are 
fixed at $\theta^{(i)} 
=\pi/4$ and $\varphi = \pi/4$.  The different curves correspond to 
different magnetic field inclination angles 
($\theta_B$ is the angle between $\bB$ and the surface normal).  As in Fig.~2, the 
light lines (labeled ``no ion'') show results when ion motion is neglected in the 
plasma dielectric 
tensor.}
\label{fig5}
\end{figure}
\begin{figure}
\epsscale{1.0}
\plotone{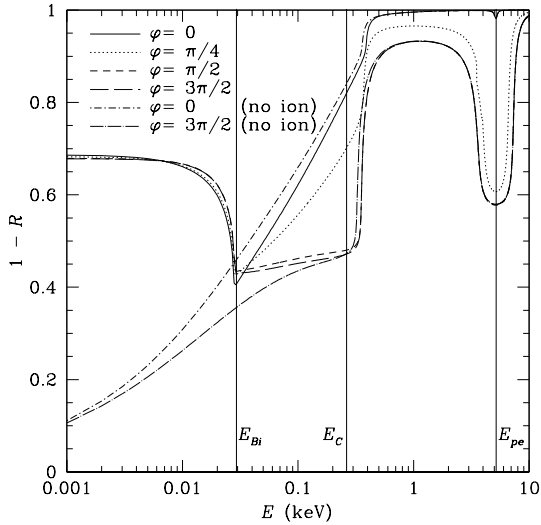}
\caption{Same as Fig.~5, except that the geometry is 
fixed at $\theta^{(i)}=\pi/4$ and $\theta_B=\pi/4$, and the different
curves correspond to different $\varphi$ (the angle of the plane 
of incidence with respect to the xz plane; see \S\ref{sect:Reflect}).}
\label{fig6}
\end{figure}
\begin{figure}
\epsscale{1.0}
\plotone{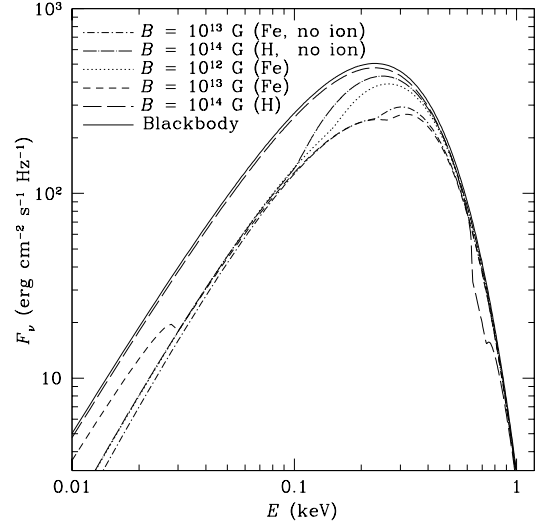}
\caption{Spectral flux as a function of 
photon energy $E$ for the cases of condensed Fe 
($B=10^{12},10^{13}$ G) and H ($B=10^{14}$ G)
surfaces, all at temperature $T = 10^6$ K. The light lines (labeled ``no ion'') 
show the flux for Fe and 
H surfaces when ion motion is neglected.
The solid line shows the 
blackbody spectrum at $10^6$ K.  
For all of the curves, the magnetic 
inclination angle $\theta_B=0$.
}
\label{fig7}
\end{figure}

When calculating the emissivity, it is clear that the inclusion of the ion terms 
in eqs.~(\ref{eq:DieEle}) for the elements of the dielectric tensor can qualitiatively 
change the emission spectrum at low energies (see Figs.~\ref{fig2}--\ref{fig7}).  
As discussed in \S 2.2, complete neglect of ion effects is not justified; while 
the exact dielectric tensor is currently unknown, the true spectra should lie 
between the two limiting cases we present here.
Without the ion terms, the broad feature around $E_{Bi}$ is replaced by a stronger 
depression of $J$ at low energies, up to $E\sim E_C$.  At high energies, the ion 
effect is unimportant.

Figures \ref{fig5} and \ref{fig6} 
give some examples of our numerical results for the cases when the magnetic field 
is not perpendicular to the surface (i.e., $\theta_B \ne 0$).  In these cases the emissivity $J$ is 
no longer symmetric with respect to the surface normal, but depends on $\theta^{(i)}$ and the azimuthal 
angle $\varphi$.  Although the geometry is more complicated, the basic features of the emissivity are 
similar to those depicted in Figs.\ \ref{fig2}--\ref{fig4}.

\begin{figure}
\epsscale{1.0}
\plotone{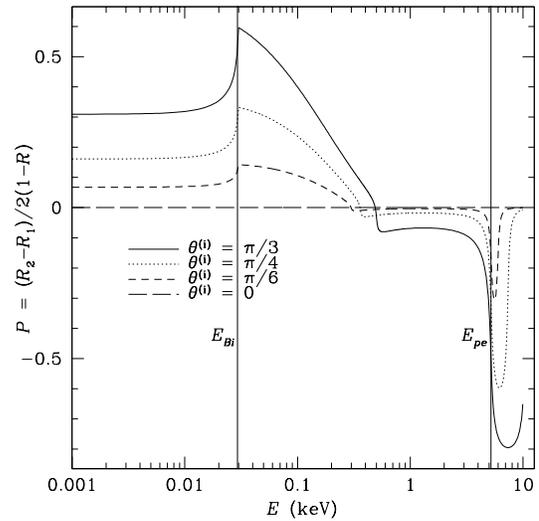}
\caption{Degree of linear polarization $P$ 
[see eq.~(\ref{eq:EmitPol})] as a function of photon energy $E$ for the 
case of condensed Fe surface, with $B=10^{13}$ G. The $B$ field is normal
to the surface, and the different curves correspond to different 
angles $\theta^{(i)}$ between incident photon direction and surface normal.
The net linear polarization is peaked around $E_{Bi}$ and $E_{pe}$.  
Positive $P$ corresponds to polarization parallel to the 
$\bk$-$\bB$ plane, while negative $P$ corresponds to polarization 
perpendicular the $\bk$-$\bB$ plane. Note that $P$ changes sign
around $E_C$.}
\label{fig8}
\end{figure}

\begin{figure}
\epsscale{1.0}
\plotone{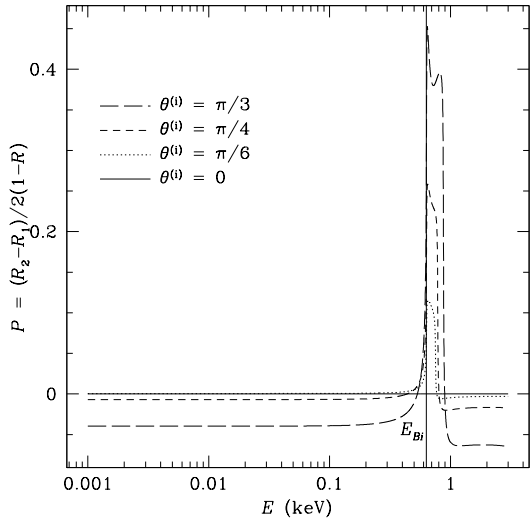}
\caption{Same as Fig.~8, except for the H surface at $B=10^{14}$~G.  
There is a slight net linear polarization perpendicular to the 
$\bk$-$\bB$ plane ($P\sim -5\%$), except around 
$E_{Bi}$ where the polarization peaks parallel to the $\bk$-$\bB$ plane.}
\label{fig9}
\end{figure}

Figure \ref{fig7} depicts specific flux at the NS surface, {$F_{\nu} = \int_0^{2\pi} d\varphi \int_0^{\pi/2} d\theta^{(i)} \cos\theta^{(i)} 
\sin\theta^{(i)} J(\theta^{(i)},
\varphi)\,B_{\nu}(T)$}, as a function of photon energy for the three 
cases illustrated in Figs.\ \ref{fig2}--\ref{fig4}.  For the Fe surface, 
there is a reduced emission (by a factor of 2 or so) around $E_{Bi}\la E\la E_c$ compared 
to the blackbody at the same temperature. For the H surface
at $B=10^{14}$~G, the flux is close to blackbody at all 
energies except for a broad feature around $E_{Bi}$.

The radiation from the condensed surface is polarized.  
Figures \ref{fig8} and \ref{fig9} show the degree
of linear polarization as a function of energy for the cases 
illustrated in Figures \ref{fig3} and \ref{fig4}
(i.e., Fe at $10^{13}$ G and H at $10^{14}$ G).  The degree of
linear polarization $P$ increases with angle of incidence, 
and is clearly peaked around $E_{Bi}$ and $E_{pe}$.  
For the Fe surface, at energies below $E_{C}$, 
the polarization vector is parallel to the $\bk$-$\bB$ plane. 
Above $E_{C}$, the sign of $P$ changes, and the radiation 
is polarized perpendicular to the $\bk$-$\bB$ plane. 
For H, there is a slight net linear polarization perpendicular 
to the $\bk$-$\bB$ plane, except near $E_{Bi}$, where the polarization
peaks with $P>0$. These polarization properties of
condensed surface emission are qualitatively different from
those of atmosphere emission
\citep[see][and references therein]{LaiHo03b}.

\subsection{Comparison with Previous Work}

\begin{figure}
\epsscale{1.0}
\plotone{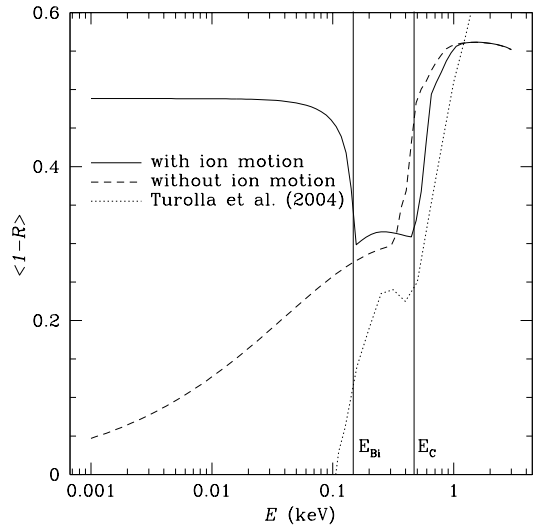}
\caption{Angle-averaged intensity $\langle 1-R \rangle$ as defined in \S 4.1 
for $B=5\times 10^{13}$ G, $T_6=1.0$, 
$\theta_B=0.7\times\pi/2$. The solid line shows our result including the
ion effect, while the dashed-line shows the results when the ion motion
is neglected. For comparison, the dotted line shows data from Fig.~5
of Turolla et al. (2004).}
\label{fig10}
\end{figure}

Recently, Turolla et al.(2004) performed a detailed calculation of 
the emissivity of a solid Fe surface. Our results differ significantly 
from theirs in several respects. In particular, Turolla et al. 
found that collisional damping in the condensed matter leads to a sharp
cut-off in the emission at low photon energies, especially when the 
magnetic field is inclined with respect to the surface normal.
For comparison, in Fig.~\ref{fig10} we show the angle-averaged emissivity,
$\langle 1-R \rangle = F_\nu/[\pi B_\nu(T)]$,
for a specific case with $B=5\times 10^{13}$~G, $T_6=1.0$ and 
$\theta_B=0.7\times\pi/2$; this should be directly compared with Fig.~5
in Turolla et al. Their results show no emission below $\sim 0.1$ keV, and
they find that this ``cutoff'' feature becomes more pronounced 
as the magnetic field inclination angle increases and the field strength 
decreases. Our calculations clearly do not show this behavior (see the
solid line of Fig.~\ref{fig10}).

These discrepancies stem from at least 
two differences in the reflectivity calculation: 
(1) Turolla et al. neglected the effect of ion motion in 
their expression for the plasma dielectric tensor (see 
the end of \S 2.2). This strongly 
affects the emissivity at $E\la E_{Bi}$ (see also Figs.~\ref{fig2}--\ref{fig7}).
(2) Even when the ion motion is neglected (by setting $m_i=\infty$),
our result (see the dashed-line in Fig.~10) does not reveal any low-energy 
cutoff. It is most likely that this difference arises from 
the ``one-mode'' description for the transmitted radiation adopted by
Turolla et al.: when the real part of the index of refraction of a mode is 
less than zero or the imaginary part of the index of refraction exceeds 
a threshold value, this mode is neglected by Turolla et al.
in the transmitted wave. Such treatment is incorrect and can lead
to significant errors in the reflectivity calculation. 
The inclusion of collisional damping gives rise to complex values for 
the index of refraction, which lead to transmitted 
waves with a propagating (oscillatory) part multiplied by a 
decaying amplitude (see Appendix B). While the damping factor for such 
waves can be large if the index of refraction has a large imaginary part, 
the propagating piece allows energy to be carried across the vacuum-surface 
boundary, therefore these waves cannot be ignored in the reflectivity 
calculation.\footnote{After submitting our paper to ApJ, a preprint 
by \cite{perez-azorin} appeared on the archive, reproducing the calculations 
described here.  They arrive at conclusions similar to the ones discussed above.}

\section{Discussion}
\label{sect:Concl}

As discussed in \S 1, many isolated NSs detected in thermal emission 
display no spectral features and are well fit by a blackbody spectrum. 
The most thoroughly studied object of this type is RX J1856.5$-$3754,
which is well fit in the X-ray by a blackbody spectrum at 
$kT_\infty=63.5$~eV, with emission radius 
$R_\infty=4.4\,(d/120~{\rm pc})$~km (where $d$ is the distance).
This X-ray blackbody underpredicts the optical flux by a factor of 7. 
\citet{PZ03} review several models involving a non-uniform 
temperature on the surface of the NS, in which the 
X-ray photons are emitted by a hot spot. By varying the temperature
distribution and assuming blackbody emission from each surface element,
a reasonable fit to the X-ray and optical data can be achieved
\citep[see also][]{BrajeRomani,Truemper-ea}.
Nevertheless, the nearly perfect X-ray blackbody spectrum of 
RX J1856.5$-$3754 is surprising.

If the NS surface is indeed in the condensed form (see \S2), the
emissivity will be determined by the properties of the condensed
matter. Our calculations (\S 3 and \S 4) show that the emission 
spectrum resembles that of a diluted blackbody, with the reduction
factor in the range of $J=0.4-1$ depending on the photon energy
(see Figs.\ \ref{fig2}--\ref{fig6}). 
This would increase the inferred emission radius by
a factor of $J^{-1/2}$. The weak ``absorption'' features in the emission
spectrum are associated ion cyclotron resonance and electron plasma frequency
in the condensed medium. We note that the emissivity and spectrum
presented in this paper correspond to a local patch of the NS.
When the emission from different surface elements are combined to form 
a synthetic spectrum, these absorption features are expected to be smoothed 
out further because of the magnetic field variation across the NS surface.

\begin{figure}
\epsscale{1.0}
\plotone{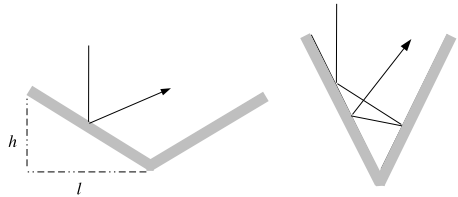}
\caption{Effect of surface roughness on the reflectivity.
The surface roughness is characterized by the vertical scale $h$ and
horizontal scale $l$, both much greater than the photon wavelength.
For the idealized ``triangular'' surface, 
a normal incident ray goes through at least two reflections if
$\theta=\tan^{-1}(l/h)<60^\circ$, at least three reflections if
$\theta<36^\circ$, at least four reflections if
$\theta<180^\circ/7$, etc. Thus net reflectivity of the rough surface
is $\ll 1$ if $h>{\rm a~few}\times l$, and the emission spectrum 
will be close to blackbody.}
\label{fig11}
\end{figure}

In our calculations, we have assumed a perfectly smooth surface.
This is valid if the condensed matter is in a liquid state, as is likely
to be the case for H condensate (see \S 2.1). For Fe, the condensed
surface is most likely a solid and we may expect a rough surface.
Although it is not possible to predict the scale and shape of the surface
irregularities, their maximum possible height $h_{\rm max}$ can be
estimated from the requirement that the stress nonuniformity
$\sim\rho g h$ is small compared to the shear stress
$\mu\theta_s$. With the shear modulus $\mu\simeq 0.1\,n_i\,(Ze)^2/a_i$
(e.g., \citet{OgataIchimaru}) and the maximum strain angle $\theta_s
=10^{-3}\theta_{-3}$, we find 
$h_{\rm max}\sim 2\times 10^{-5}\,\theta_{-3}\,
Z^2 A^{-4/3}\rho_1^{1/3}g_{14}^{-1}$~cm
(where $g=10^{14}g_{14}$~cm~s$^{-2}$ is the surface gravity).
For the condensed Fe surface at the density given by 
eq.~(\ref{rho_s}), we have $h_\mathrm{max}\sim
4\times 10^{-4}\,\theta_{-3}\, B_{12}^{2/5}$~cm (for a neutron star with $R=10$~km
and $M=1.4\,M_\odot$).
Clearly, the scale of the surface roughness can easily be much larger 
than the photon wavelength ($\sim 10$\AA). As illustrated in Fig.~\ref{fig11},
the surface may be much less reflective than 
the results shown in \S 4, and the emission will be closer to the blackbody
spectrum.

The emission from a condensed NS surface is distinct from atmospheric
emission in several aspects: (i) Atmospheric emission
generally possesses a hard spectral tail (although this tail is 
somewhat suppressed by the QED effect for $B\ga 10^{14}$~G; see \S1), 
whereas the condensed surface emission does not;
(ii) The spectrum of a cool NS atmosphere can have both cyclotron and
atomic absorption features (again, they are reduced for 
$B\ga 10^{14}$~G); the broad (cyclotron and plasma) features of 
condensed surface emission persist even in the magnetar field regime
(if they are not smoothed out by variations of surface B fields or 
by the rough surface effect);
(iii) The polarization signature of condensed matter emission 
is qualitatively different from that of atmospheric emission.
All these aspects can serve as diagnostics for the physical condition 
of the emission region.

At the surface temperature of AXPs and SGRs, $T\simeq 5\times 10^6$~K, 
H is unlikely to be condensed, but Fe condensation is possible.
The dim, isolated NSs have lower temperatures ($T\la 10^6$~K),
and if they possess magnetar-like fields, condensation is likely.
In particular, the black-body X-ray spectra of RX J1856.5$-$3754 ($kT\simeq 64$eV) 
and RX J0420$-$5022 ($kT\simeq 45$eV; see \citealt{Haberl-ea04b})
could arise from condensed surface emission (e.g., 
non-smooth Fe surface at $B\ga 10^{12}$~G), although to
account for the optical data, nonuniform surface temperatures are still 
needed.

\acknowledgements
D.L. thanks Richard Epstein for a conversation on reflection from
non-smooth surfaces (summer 1998).
A.P. acknowledges the hospitality
of the Astronomy Department of Cornell University.
We would like to thank the anonymous referee for his/her 
comments and suggestions.
This work is supported in part by 
NSF grant AST 0307252, NASA grant NAG 5-12034 and SAO grant TM4-5002X.
The work of A.P. is supported in part by
RFBR grants 05-02-16245 and 03-07-90200
and by the ``Russian Leading Scientific Schools'' grant 1115.2003.2.

\appendix
\section{A. Reflectivity Calculation}
\label{sect:A}

Here we fill in some of the details for the reflectivity 
calculation described in \S\ref{sect:Reflect}. 

In the coordinate system xyz defined in \S\ref{sect:Reflect}, the explicit expression for eq.~(\ref{eq:ModeEq}) is:
\begin{eqnarray}
\label{eq:ExpLambda}
\left[\begin{array}{cc}
\epsilon\cnbs+\eta\snbs + n_j^2(\snjs\cnps-1) & i\,g\cnb + n_j^2\snjs\snp\,\cnp\\ 
-i\,g\cnb+n_j^2\snjs\snp\cnp & \epsilon+n_j^2(\snjs\snps-1)\\
(\epsilon-\eta)\snb\cnb+n_j^2\snj\cnj\cnp & i\,g\,\snb+n_j^2\snj\cnj\snp\\
\end{array}\right.\nonumber\\
\left.\begin{array}{c}
(\epsilon-\eta)\snb\cnb+n_j^2\snj\cnj\cnp\\
-i\,g\snb+n_j^2\snj\cnj\snp\\
\epsilon\snbs+\eta\cnbs-n_j^2\snjs
\end{array}\right]\cdot \left[
\begin{array}{c}
E_x\\
E_y\\
E_z\end{array}\right]\ =\ {\bf 0},
\end{eqnarray}
where $n_j$ (with $j=1,2$) is the index of refraction in the medium,  
and $\theta_j^{(t)}$ is the 
formal complex angle 
of propagation calculated using Snell's law  (see Appendix B for
a discussion of the interpretation of complex $\theta^{(t)}_j$). 
Taking the determinant of eq.~(\ref{eq:ExpLambda}) yields:
\be
\label{eq:DetLambda}
a_4 n_j^4 + a_2 n_j^2 + \cnj\sni\left(a_1 n_j +
a_3 n^3_j\right) + a_0 = 0,
\ee
where we have used Snell's law and the following definitions:
\begin{subequations}
\begin{eqnarray}
\label{eq:DetDefs}
a_0 & = & (\epsilon^2-g^2)\eta+{1\over 8}\left[g^2+\epsilon(\eta-\epsilon)\right]\left(2+
6\cntb-4\snbs\cntp\right)\snis\nonumber\\
& & -2\left(\epsilon\cnbs+\eta\snbs\right)\snps\snif\\
a_1 & = & \left[\epsilon(\eta-\epsilon)+g^2\right]\sntb\cnp\\
a_2 & = & {1\over 2}\left[g^2-\epsilon(\epsilon+3\eta)-(g^2+\epsilon(\eta-\epsilon))\cntb\right]+
\nonumber \\
& & \left[\epsilon(\epsilon-\eta)\cntb+(\epsilon\cnbs+\eta\snbs)\snps-\epsilon\cnps\right]\snis\\
a_3 & = & (\epsilon-\eta)\sntb\cnp\\
a_4 & = & \left(\epsilon^2-g^2\right)\eta.
\end{eqnarray}
\end{subequations}\hspace{-4pt}
The $\cnj$ term is moved to the right-hand side, and the entire equation 
is then squared. Using the identity $\cnjs=1-\snjs$ and Snell's law 
gives a polynomial equation in $n_j$:
\begin{eqnarray}
\label{eq:PolyEq}
a_4^2 n_j^8 + (2 a_2 a_4 - a_3^2\snis) n_j^6 + (a_2^2+ 2 a_0 a_4 -2 a_1 a_3\snis + a_3^2\snif) n_j^4 
+ \nonumber\\
(2 a_0 a_2 - a_1^2\snis + 2 a_1 a_3\snif) n_j^2 + a_0^2 +a_1^2\snif\ =\ 0
\end{eqnarray}
The polynomial equation (\ref{eq:PolyEq}) has eight roots for $n_j$ which we found numerically 
using Laguerre's method (Press et al, 1996).  Only two of the roots are physical and satisfy the 
original 
equation (\ref{eq:DetLambda}).  In practice, it was found that for certain combinations of 
the parameters $E$, $\theta^{(i)}$, 
$\theta_B$, $\varphi$, a spurious root satisfies eqn.~(\ref{eq:DetLambda}) to the specified degree of 
accuracy, 
resulting in an unphysical result for the reflectivity.  It is often the case that such roots can be 
discounted physically using the conditions (\ref{eq:Cond1}) and (\ref{eq:Cond2}) (see Appendix B). 
Once the indices of refraction $n_1$, $n_2$ are known, the normal mode polarization vectors can be 
determined.  Solving eqn.~(\ref{eq:ModeEq}) for the ratios $E^{(t)}_x/E^{(t)}_y$ and 
$E^{(t)}_z/E^{(t)}_y$ 
gives the expressions:
\begin{subequations}
\begin{eqnarray}
f_j=\left({E^{(t)}_x\over E^{(t)}_y}\right)_j & = & 
i\,{{\epsilon-i\,B_j\,g\snb+n_j^2\snj\snp\left(B_j\cnj+\snj\snp\right)-n_j^2}
\over {g\cnb+A_j\,g\snb+i\,n_j^2\snj\left(A_j\cnj+\snj\cnp\right)\snp}} ,
\\
g_j=\left({E^{(t)}_z\over E^{(t)}_y}\right)_j & = & A_j\left(E^{(t)}_x\over E^{(t)}_y\right)_j+B_j ,
\\
A_j & = & -{{\epsilon\cnbs+\eta\snbs+n_j^2\snjs\cnps}\over {(\epsilon-\eta)\snb\cnb+n_j^2\snj\cnj\cnp}} ,
\\
B_j & = & -{i\,g\cnb+n_j^2\snj\snp\cnp\over (\epsilon-\eta)\snb\cnb+n_j^2\snj\cnj\cnp} .
\end{eqnarray}
\label{eqn:E/E}
\end{subequations}\hspace{-4pt}
With the propagation modes in the plasma determined, the latter two equations of (\ref{eq:BCs}) give:
\begin{equation}
\label{eq:ExpBCs}
\left(\begin{array}{cccc}
\cni\snp & \cnp & C_1 & C_2\\ 
\cni\cnp & \snp & -C_1 & -C_2\\
-\cnp & -\cni\snp & C_5 & C_6\\
-\snp & \cni\cnp & C_7 & C_8 
\end{array}\right)\cdot
\left\{\begin{array}{c}
\left(\begin{array}{c}
r_{11}\\r_{12}\\t_{11}\\t_{12}\end{array}\right)\\
\left(\begin{array}{c}
r_{21}\\r_{22}\\t_{21}\\t_{22}\end{array}\right)
\end{array}\right. = 
\left.\begin{array}{c}
\left(\begin{array}{c}
-\cni\snp\\\cni\cnp\\\cnp\\\snp
\end{array}\right)\\
\left(\begin{array}{c}
-\cnp\\-\snp\\-\cni\snp\\\cni\cnp
\end{array}\right)
\end{array}\right.
\end{equation}
for the incoming polarization modes $\bfe^{(i)}_1 = (-\cni\cnp,-\cni\snp,\sni)$ and $\bfe^{(i)}_2 = 
(\snp,-\cnp,0)$.  Inverting the coefficient matrix of eqn.~(\ref{eq:ExpBCs}) and performing extensive 
algebra yields the following expressions for the reflected field amplitudes:
\begin{subequations}
\begin{eqnarray}
r_{11} & = & {{4 A\cni-2 B_-\snis+(3+\cnti)(B_+\cntp+C_+\sntp)}\over {4 A\cni + B_-(3+\cnti) - 
2\snis(B_+\cntp+C_+\sntp)}} ,
\\
r_{12} & = & {{4\cni(C_-+B_+\sntp-C_+\cntp)}\over {4 A\cni + B_-(3+\cnti) - 
2\snis(B_+\cntp+C_+\sntp)}} ,
\\
r_{21} & = & {{4\cni(C_+\cntp-B_+\sntp+C_-)}\over {4 A\cni + B_-(3+\cnti) - 
2\snis(B_+\cntp+C_+\sntp)}} ,
\\
r_{22} & = & {{(3+\cnti)(B_+\cntp+C_+\sntp)-4 A\cni-2 B_-\snis}\over {4 A\cni + B_-(3+\cnti) - 
2\snis(B_+\cntp+C_+\sntp)}} ,
\end{eqnarray}
\end{subequations}\hspace{-4pt}
using the definitions:
\begin{subequations}
\begin{eqnarray}
C_{1,2} & = & \left[1+\left|f_{1,2}\right|^2+\left|g_{1,2}\right|^2\right]^{-1/2} , \\
C_{5,6} & = & C_{1,2}\left[n_{1,2}\,\sin\theta^{(t)}_{1,2}\,\cos\varphi\,g_{1,2}-n_{1,2}\,\cos\theta^{(t)}_{1,2}\,f_{1,2}\right] , \\
C_{7,8} & = & C_{1,2}\left[n_{1,2}\,\sin\theta_{1,2}^{(t)}\,\sin\varphi\,g_{1,2}-n_{1,2}\,\cos\theta_{1,2}^{(t)}\right] , \\
A & = & (C_6 C_7-C_5 C_8) , \\
B_{\pm} & = & (C_2 C_5+C_2 C_6)\pm(C_2 C_7-C_1 C_8) , \\
C_{\pm} & = & (C_1 C_6+C_2 C_5)\pm(C_2 C_7-C_1 C_8) ,
\end{eqnarray}
\end{subequations}\hspace{-4pt}
The reflectivity and the emission spectrum and polarization are then determined by eqs. 
(\ref{eq:Spectrum}), 
(\ref{eq:TotSpec}), and (\ref{eq:EmitPol}).

\section{B. Complex Angle of Propagation}

In this Appendix we outline some of the physical properties of the wave 
propagating in the plasma with complex index of refraction
(cf. Born \& Wolf 1970, \S 13.2).

For a medium with complex index of refraction $n=n_R+i n_I$ (where $n_R$ and $n_I$ are real), the formal 
refraction angle $\theta^{(t)}$, as determined by Snell's law, is complex.
Let $\cn=(1-\sin^2\theta^{(t)})^{1/2}=\cnre+i\cnim$.  Defining 
the vector parallel to the plane of incidence $\hat\bfs=(-\cnp,-\snp,0)$, 
the wavevector for the transmitted waves can be written:
\begin{eqnarray}
\label{eq:WavVec}
\bk^{(t)} & = & {n\omega\over c}\left(\sn\hat\bfs-\cn\hatz\right) \nonumber\\
& = & {\omega\over c}\left[\sni\hat\bfs+(n_I\cnim-n_R\cnre)\hatz-i(n_I\cnre+n_R\cnim)\hatz\right]
\end{eqnarray}
The transmitted electric field has the form $E^{(t)}\propto e^{i\bk^{(t)}\cdot{\bf r}-i\omega t}$.  
Substituting eqn.~(\ref{eq:WavVec}) into this expression, the field takes the form:
\begin{eqnarray}
\label{eq:Etrans}
E^{(t)} & \propto & \exp\left[(n_R\cnim+n_I\cnre)z\right]\times\nonumber\\
& & \exp\left[i{\omega\over c}\left(\sni s-(n_R\cnre-n_I\cnim)z\right)-i\omega t\right].
\end{eqnarray}
Thus, the transmitted wave has a propagating component multiplied by a damping factor.  Since the amplitude 
of the wave must decrease as it travels through the medium, eqn.~(\ref{eq:Etrans}) gives the following 
condition on the index of refraction (recall that in the geometry of \S\ref{sect:Reflect}, $z<0$):
\be
\label{eq:Cond1}
n_R\cnim+n_I\cnre>0.
\ee
The traveling component can be used to define a new wavevector 
$\bk' = \sni\hat\bfs - (n_R\cnre-n_I\cnim)\hatz$.
The real angle of propagation is then given by 
$\cos\theta^{(t)'}=\hat{\bk'}\cdot\bk'/|\bk'|$.
By assumption, the angle of propagation for the refracted 
wave measured with respect to the $z$ axis must be greater 
than $\pi/2$, yielding a second condition on the index of refraction:
\be
\label{eq:Cond2}
-1\le
\cos\theta^{(t)'}=
{{n_I\cnim-n_R\cnre}\over {\sqrt{\snis+(n_I\cnim-n_r\cnre)^2}}}\le 0.
\ee
The real and imaginary parts of the indices of refraction for the birefringent transmitted waves must 
satisfy eqs. (\ref{eq:Cond1}) and (\ref{eq:Cond2}).



\end{document}